\documentclass[journal,comsoc,11pt]{IEEEtran}
\IEEEoverridecommandlockouts
\usepackage{cite}
\usepackage{graphicx}
\usepackage{amsmath}
\usepackage{amssymb}
\usepackage{algorithmicx}
\usepackage{algorithm}
\usepackage{color}
\usepackage{bm}
\usepackage{setspace}

\allowdisplaybreaks

\begin{document}

\title{
Multi-domain Cooperative SLAM: The Enabler for Integrated Sensing and Communications}

\author{
Jie Yang, Chao-Kai Wen, Xi Yang, Jing Xu, Tao Du, and Shi Jin 
\thanks{Jie~Yang is with the School of Automation and the Frontiers Science Center for Mobile Information Communication and Security, Southeast University, Nanjing, China.
Shi~Jin (corresponding author) is with the National Mobile Communications Research Laboratory and the Frontiers Science Center for Mobile Information Communication and Security, Southeast University, Nanjing, China.
Xi~Yang, Jing~Xu, and Tao~Du are with the National Mobile Communications Research Laboratory, Southeast University, Nanjing, China. 
Chao-Kai~Wen is with the Institute of Communications Engineering, National Sun Yat-sen University, Kaohsiung, Taiwan.
}


}

\maketitle	

\begin{abstract}
	\setlength{\baselineskip}{10.5pt}Simultaneous localization and mapping (SLAM) provides user tracking and environmental mapping capabilities, enabling communication systems to gain situational awareness. Advanced communication networks with ultra-wideband, multiple antennas, and a large number of connections present opportunities for deep integration of sensing and communications. First,  the development of integrated sensing and communications (ISAC) is reviewed in this study, and  the differences between ISAC and traditional communication are revealed. Then, efficient mechanisms for multi-domain collaborative SLAM are presented. In particular, research opportunities and challenges for cross-sensing, cross-user, cross-frequency, and cross-device SLAM mechanisms are proposed. In addition, SLAM-aided communication strategies are explicitly discussed. We prove that the multi-domain cooperative SLAM mechanisms based on hybrid sensing and crowdsourcing can considerably improve the accuracy of localization and mapping in complex multipath propagation environments through numerical analysis. Furthermore, we conduct testbed experiments to show that the proposed SLAM mechanisms can achieve decimeter-level localization and mapping accuracy in practical scenarios, thereby proving the application prospect of multi-domain collaborative SLAM in ISAC.      
\end{abstract}

\section{Introduction}
\begin{figure*}
	\vspace{-0.7cm}
	\centering
	\includegraphics[scale=0.46]{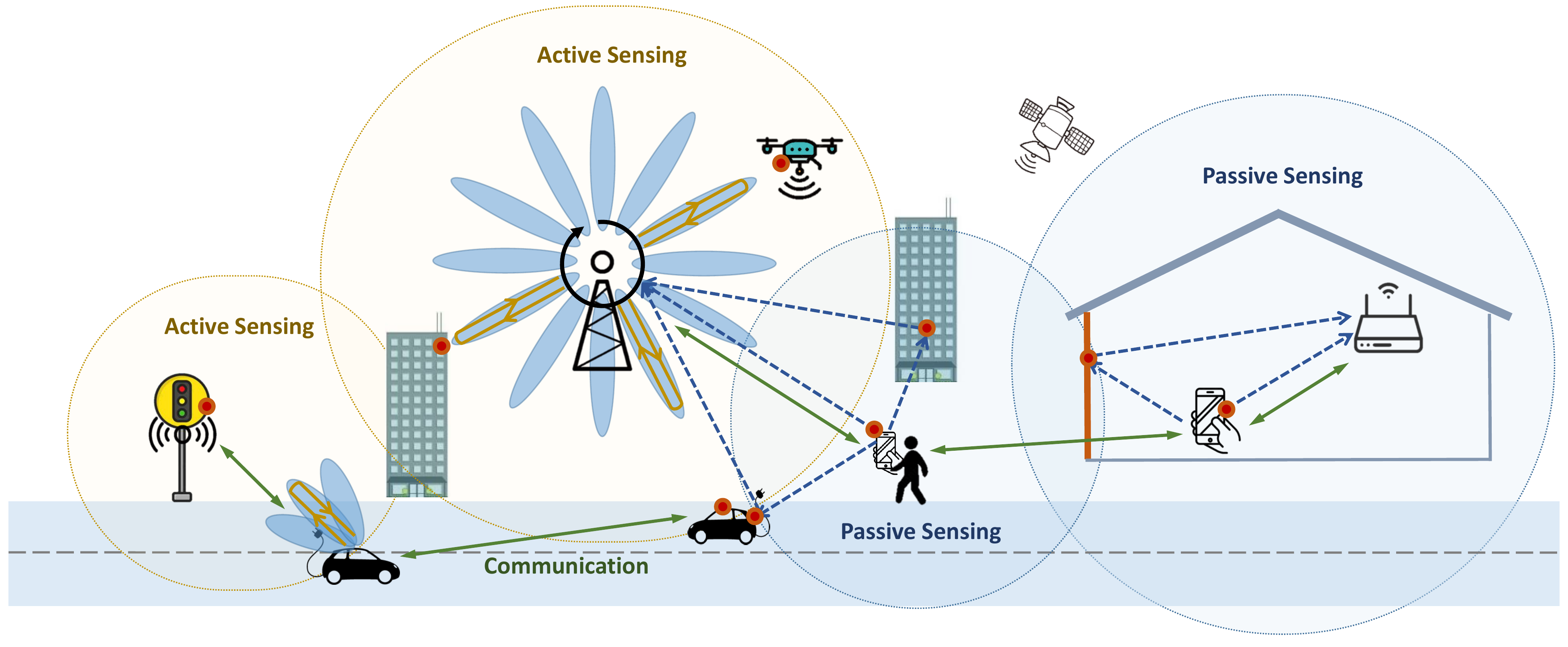}
	\caption{Illustration of multi-domain cooperative ISAC networks, where the ISAC system should provide services for many users, including vehicles, mobile phones, and IoT devices. The ISAC system interacts with GNSS, UAV, radar, WLAN, and sensor networks. Cooperation among various sensing types and networks is imperative.}
	\label{system}\vspace{-0.2cm}
\end{figure*}

Future wireless communication systems are expected to provide ubiquitous connectivity with ultrahigh throughput and reliability and ultralow latency; they are also expected to realize the ability to sense, control, and even optimize wireless environments \cite{6G0}. New applications, such as extended reality, holographic communications, autonomous driving, smart healthcare, and intelligent industry, have emerged. They require mass data transmission, centimeter-level localization, and high fine-grained environmental information with the rapid advent of the intelligent age. Integrated sensing and communications (ISAC) is anticipated to play a pivotal role in achieving these applications\cite{ISAC00}.
 
Continuing developments have made wireless communications capable of sensing. On the one hand, exploited millimeter wave (mmWave) and terahertz (THz) frequencies coincide with the spectrum of mmWave radar and high-resolution THz imaging radar. Large bandwidths result in high range resolutions. On the other hand, upscaled antenna deployment enables unprecedented angular resolution. Future wireless communication systems coupled with ultradense network deployment enable a paradigm shift in sensing capabilities. Location information can aid in addressing several key challenges in communication systems, including increases in traffic and number of devices, robustness for mission-critical services, and reduction in total energy consumption and latency \cite{loc}. A perceptive mobile network is proposed to integrate sensing into communications and share the majority of system modules and the same transmitted signals \cite{ISAC2}. Existing studies on the fundamental limits of ISAC are comprehensively reviewed in \cite{ISAC11}. Four categories of ISAC use cases, including high-accuracy localization and tracking, simultaneous localization and mapping (SLAM), augmented human sense, and gesture and activity recognition, are highlighted in \cite{ISAC1}. 

Although considerable advantages of ISAC have been predicted, deep integration of sensing and communications still requires further investigation. SLAM can provide locations of user equipment (UE) and radio features in the propagation environment. These results help design sensing-aided communication strategies. Therefore, SLAM is a promising technology for achieving deep integration of sensing and communications. However, SLAM faces critical challenges in communication systems due to the massive connections and complex multipath propagation environments \cite{slam,slam2}. None of the existing works have realized the SLAM function under the 5G New Radio (NR) standard because of specification and hardware constraints. SLAM has relatively mature applications in the field of robotics; it is often achieved by leveraging the robot's sensors, such as inertial measurement unit (IMU), camera, and laser, which provide more landmarks than features available in typical communication networks \cite{slam3}. Although the resolution of a radio is lower than that of a camera or light detection and ranging (LiDAR), a radio can cover a long detection range and is vaguely affected by weather and light conditions, which are the key challenges faced by visual and LiDAR SLAM.
Moreover, completing the complex task of realizing high-quality communication and high-accuracy sensing is difficult for a single network, particularly for a single user \cite{co0}. Radio frequency (RF) signals designed for communication can help realize communication and sensing by completely reusing the communication hardware. Therefore, other devices in the communication network can participate in collaborative SLAM.
We aim to develop multi-domain cooperative SLAM mechanisms in the present study with the cooperation of multiple sensing types, users, frequency bands, and devices.

The rest of this paper is organized as follows. After briefly introducing the differences between ISAC and traditional communications, we discuss open issues and technical challenges from the perspectives of SLAM result representation, multi-domain cooperative SLAM mechanisms, and SLAM-aided communication strategies. Therefore, deep integration of sensing and communication is formed by SLAM. Then, case studies and experiments on the proposed SLAM mechanisms are conducted, and conclusions are drawn.

\section{Differences between ISAC and traditional communications}

Traditional communication systems are designed to offer high data rates and reliable connections. However, the basic idea of ISAC is to realize dual functions. On the one hand, communication systems obtain sensing functions, which can identify objects, localize targets, track devices, and map radio environments. On the other hand, communication performance in the aspects of beam management (BM), resource allocation (RA), and signal processing (SP) is enhanced through sensing. The present study focuses on integrating sensing into the existing communication-only cellular systems, among which SLAM is a unique technology that helps achieve deep integration of sensing and communication. In this section, we reveal the similarities and differences between ISAC systems and traditional communication systems. The important role of SLAM in the ISAC systems is also emphasized.
 
\begin{figure*}
	\vspace{-0.7cm}
	\centering
	\includegraphics[scale=0.42]{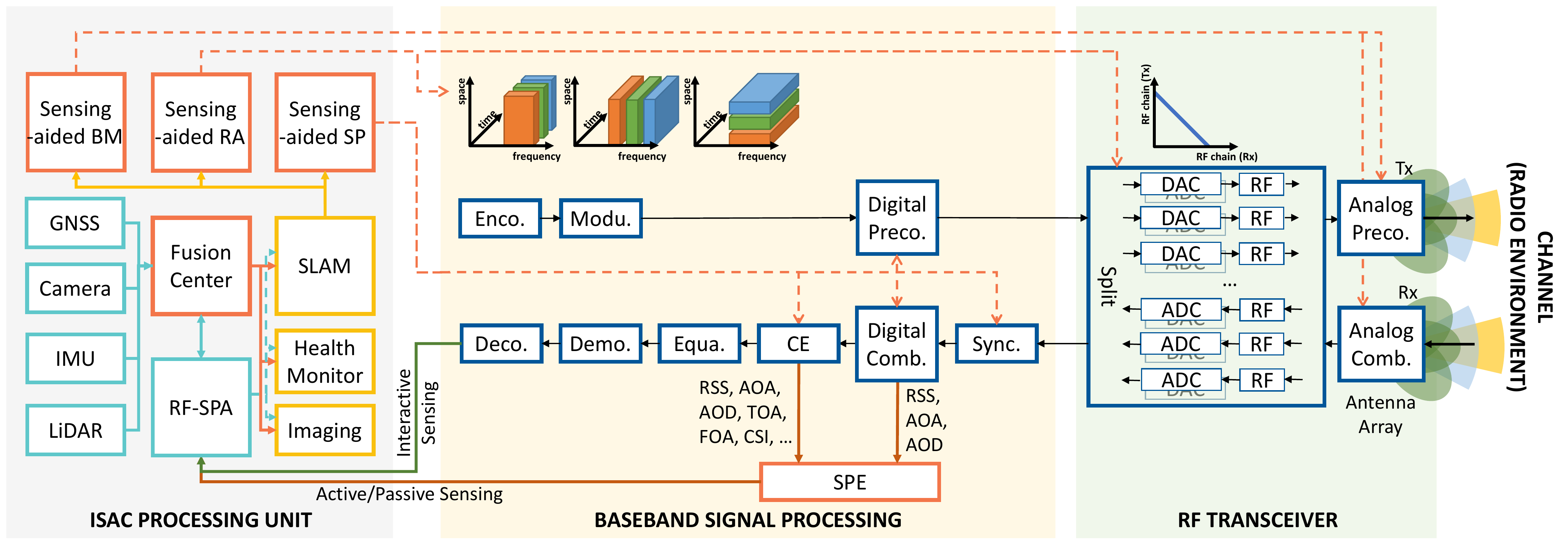}
	\caption{Architecture of a general ISAC system, including ISAC processing unit, baseband SP, and RF transceivers. Note: Enco., encoding; Deco., decoding; Modu., modulating; Demo., demodulating; Equa., equalization; Preco., precoding; Comb., combining; Sync., synchronization.}
	\label{BD}\vspace{-0.2cm}
\end{figure*}

\subsection{Multi-domain Cooperative ISAC Networks}
In the era of the Internet of Everything, communication systems should provide services for various users, including vehicles, mobile phones, and Internet of Things (IoT) devices, as shown in Fig. \ref{system}. 
Some of these users have certain sensing capabilities and can become sensing service providers.
In addition, the ISAC system is expected to interact with global navigation satellite systems (GNSSs), unmanned aerial vehicles (UAVs), radars, wireless local area networks (WLANs), and sensor networks.
Each kind of network has its advantages and disadvantages. For example, a GNSS can provide position information outdoors but is ineffective indoors; a WLAN can be economically deployed indoors, but its sensing resolution is limited. Therefore, cooperation among heterogeneous networks, which is the foundation of the multi-domain cooperative SLAM, is imperative.
 
\subsection{ISAC Protocols}
Sensing is gradually supported by default communication frame structures and protocols according to the 3GPP service and system aspect specifications.
UE and base stations (BSs) transmit uplink sounding reference signals (UL-SRS) and downlink positioning reference signals (DL-PRS), respectively. They also capture targets from echo signals and serve as conventional monostatic radars in the type of active sensing. In passive sensing, the UE or BS does not send sensing signals while capturing targets through the received signals, which are sent or reflected from targets.
The location management function enables necessary positioning information and measurement exchange among NR BSs, long-term evolution BSs, and UE in 5G core networks \cite{beam}. 
Given that sensing covers a wide range of services, where positioning is only one of them,
the protocols should be further extended to accommodate new sensing applications, such as health monitoring, imaging, and SLAM.
Sensing assistance leads to the modifications of traditional communication protocols, particularly for SLAM-aided communication protocols.
For example, a traditional exhaustive beam searching protocol requires a large number of pilots and frequent uplink feedback to find the optimal beam pair.
Overhead and latency can be greatly reduced by narrowing down the beam search space and reducing the beam sweeping period with the SLAM-aided BM protocols.

\subsection{ISAC Processing Unit}\label{IPU}
The architecture of a general ISAC system is presented in Fig. \ref{BD}, according to the perceptive mobile network proposed in \cite{ISAC2}. 
Unlike the traditional transceiver architecture, the ISAC processing unit is a specific module of the ISAC system and can be deployed in the UE, BS, or network center.
First, the ISAC processing unit needs to support multi-domain data sources, including data from RF signal, GNSS, camera, LiDAR, and IMU.
Among them, RF sensing parameter analysis (SPA) is a specialized module that works alone to provide sensing services. 
The fusion center can integrate the sensing results of different kinds of UE or devices and provide comprehensive sensing services.
The ISAC processing unit provides different sensing services, including SLAM, health monitoring, and imaging, according to the requirements.
SLAM can provide the locations of UE and radio features in the propagation environment, including the locations of BSs and reflectors. These results help design sensing-aided communication strategies, including BM, RA, and SP. Therefore, SLAM is a promising technology for achieving deep integration of sensing and communication.

\subsection{Baseband Signal Processing}\label{BB}
The baseband SP of the ISAC system is different from that of the traditional communication system. 
SLAM can play a role in baseband SP, as SLAM provides support for sensing-aided BM, RA, and SP.
Moreover, sensing-aided BM, RA, and SP assist the baseband SP modules.
First, RA is controlled by the sensing-aided RA module in the ISAC processing unit. Second, digital precoding/combining, channel estimation, and synchronization are optimized by the sensing-aided SP module. 
In addition, the sensing parameter extraction (SPE) module controls the extraction of the received signal strength (RSS), angle of arrival (AOA), angle of departure (AOD), time of arrival (TOA), frequency of arrival (FOA), or directly outputs channel state information (CSI). RF sensing parameters are obtained by active/passive sensing and then delivered to the RF-SPA module in the ISAC processing unit. The received information carrying sensing content is sent to the ISAC processing unit after decoding. This process belongs to the type of interactive sensing.

\begin{figure*}
	\vspace{-0.7cm}
	\centering
	\includegraphics[scale = 0.5]{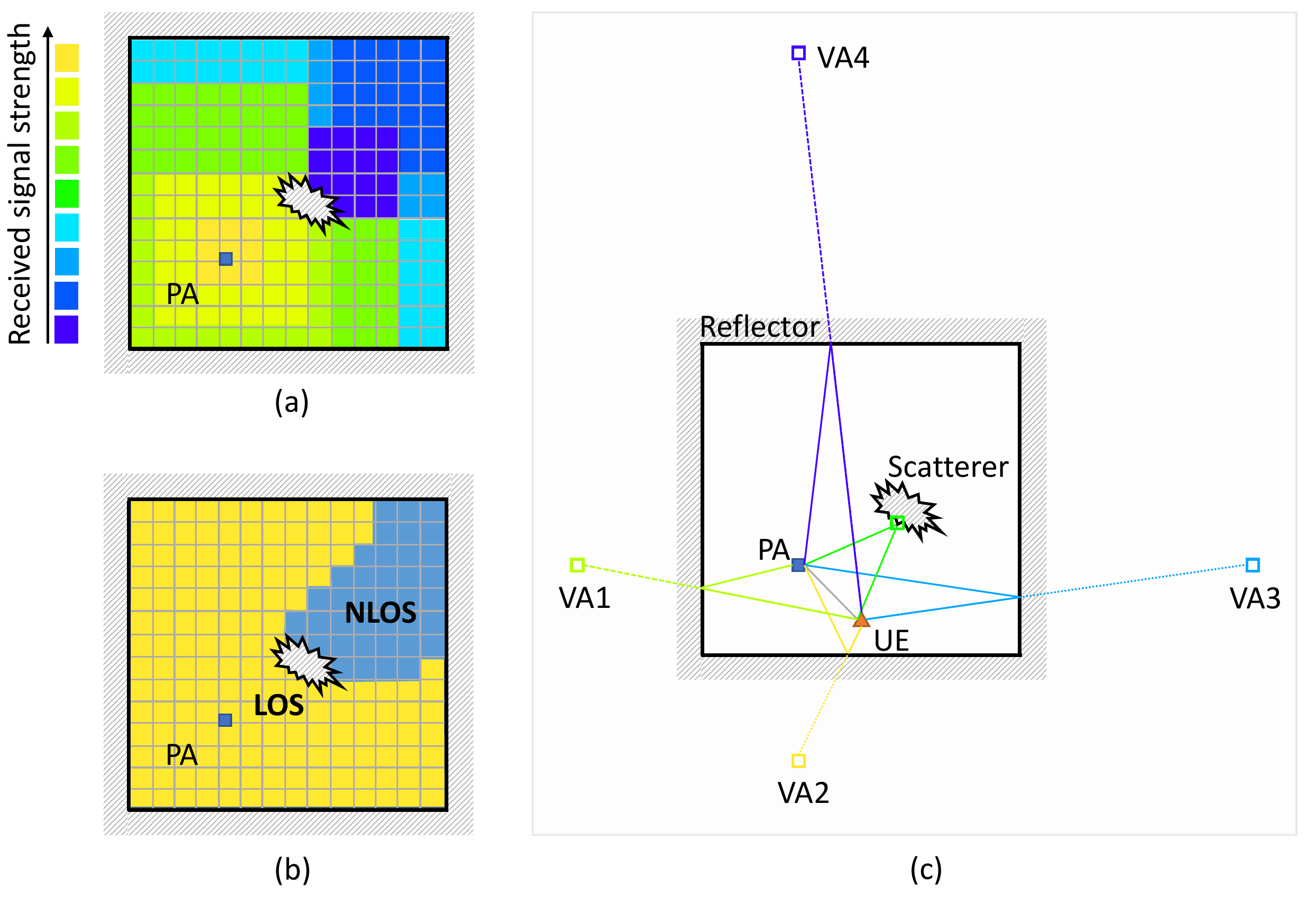}
	\caption{Illustration of typical radio maps. (a) RSS-based radio map. (b) Communication-metric-based radio map. (c) Geometric-feature-based radio map. }
	\label{RM}\vspace{-0.2cm}
\end{figure*}

\subsection{RF Transceiver}\label{RT}
The RF transceiver of the ISAC system demonstrates a split function. The number of RF chains can be split into two for simultaneous transmission and reception. Moreover, proper transmission and reception antenna isolation design must be considered.
Although the total number of RF chains is fixed, the number of RF chains for simultaneous transmission and reception can be controlled by the sensing-aided RA module. 
Analog precoding/combining can be optimized through the sensing-aided BM module. SLAM can assist the sensing-aided BM module, thereby helping the RF transceiver design. The RF transceiver of the ISAC system degenerates into that of the traditional communication systems without modifications when sensing shares the communication reference signals completely. However, the ISAC transceiver can be switched to the mode where specific sensing reference signals, beams, or RF chains are required.

\section{Open Issues and Technical Challenges}\label{cooperation}
The fusion center and sensing-aided communication modules in the ISAC processing unit are linked by SLAM techniques.
First, we introduce the radio map, which is the result representation of SLAM.
Then, we focus on the open issues and technical challenges related to communication-enabled cooperative SLAM mechanisms and SLAM-aided communication strategies.

\subsection{SLAM and Radio Map}\label{radiomap}
Radio maps visualize the RF characteristics of the physical space. 
SLAM technology can play an important role in radio map construction.
Therefore, we first discuss the representations of radio maps.
Uniform and effective representation of the radio maps is vital for ISAC systems because (1) information from different sources can be fused via the same radio map format, and (2) the radio map can provide guidance for different kinds of UE. Fig. \ref{RM} presents several typical radio maps. Region of interest (ROI) is gridded. The RSS at each grid is collected and recorded in the RSS-based radio map, also known as the RF fingerprint (Fig. \ref{RM}[a]). 
Communication-metric-based radio map \cite{UAV} also requires an offline collection process, where we turn the collected RSS or CSI into communication metrics through proper calculations, such as channel capacity, outage probability, and line-of-sight (LOS) and non-LOS (NLOS) indicators (Fig. \ref{RM}[b]). The geometric-feature-based radio map (Fig. \ref{RM}[c]) differs from the first two radio maps because griding the ROI and measuring each grid point are unnecessary. Radio features are abstracted into several static and sparse points, such as physical anchors (PAs; a PA can be a BS), virtual anchors (VAs; mirrors of PAs that represent specular reflectors), and scatterers. Therefore, a small amount of data is required to describe the geometric-feature-based radio map, which can intuitively provide guidance to beam direction prediction.
SLAM mainly exploits the location and state invariance of radio features in the propagation environment.
Thus, geometric-feature-based radio maps can be easily constructed from measurements at successive moments by SLAM techniques.

\subsection{Multi-domain Cooperative SLAM}\label{MC-SLAM}
We present novel communication-enabled SLAM mechanisms with multi-domain cooperation, including cooperation through active and passive sensing, multiple UE, multifrequency bands, and multiple devices.
Key technologies of SLAM, such as data association, data fusion, and feature detection, have changed in different cooperative mechanisms. Specific SLAM algorithms \cite{slam,slam2,slam3,hap,coj}, such as belief-propagation-based SLAM, probability hypothesis density-based SLAM, and expectation propagation-based SLAM, are not summarized in this article because of the page limitation.

\subsubsection{Cross-Sensing SLAM}\label{active_passive}
Current communication systems consist of beam sweeping, channel estimation, and data transmission stages \cite{beam}. 
Different sensing types can be realized during communication processes by using communication waveforms.
Active sensing, which can be performed without establishing a connection in the communication network, has a known reference signal.
However, self-interference is severe for active sensing in communication systems without a dedicated self-interference
elimination design. Moreover, active sensing has a smaller scope than passive sensing, particularly on the UE side.
The reason is that UE has limited power, and active sensing relies on LOS paths. 
Although passive sensing requires communication link establishment and channel estimation, multipoint cooperation can be realized, thereby enhancing passive sensing by information fusion.
In summary, active and passive sensing have advantages and disadvantages. The existing studies typically ignore the cooperation of different sensing types.

Given that the location and state of the radio feature sensed by active and passive sensing are strongly related under the same environment, different sensing types can cooperate through appropriate resource configuration and unified sensing result representation (such as the geometric-feature-based radio map).
For example, the surrounding scatterers around the BS can be estimated through active sensing in the beam sweeping stage.
Then, the SLAM results of active sensing are transformed into a unified form (geometric- feature-based radio map).
Therefore, the initialization of radio features in the propagation environment is realized with the assistance of active sensing.
In the channel sounding stage, the UE may transmit pilot signals in several beam directions. The BS can execute SLAM to construct the geometric- feature-based radio map with estimated channel parameters by passive sensing.
Thus, soft information fusion for the same radio feature can be realized.
This process is one of key techniques in SLAM.
Legacy feature refinements and new feature detection are achieved by SLAM with through passive sensing.

The cooperation between active and passive sensing involves many unsolved problems.
First, self-interference is a crucial problem that plagues active sensing. Antenna isolation and multiarray extrapolation technologies are worth further research. Second, diffuse and specular reflections bring differences to theoretical modeling and data association methods. 
Therefore, data cleaning and classification techniques based on deep learning should be studied.
Moreover, practical limitations, such as clock and orientation biases, should be the focus of future investigations.

\subsubsection{Cross-User SLAM}\label{cu}
Human and machine connectivity continues to increase in the age of the Internet of Everything. Thus, the terminals are expected to share the information for cooperation. 
As discussed in Section~\ref{active_passive}, each UE can sense the local radio environment in the communication processes.
Moreover, the local radio features can be mapped by active and passive sensing cooperation. 
However, the potential benefits provided by multiple terminals are not fully utilized in the conventional system. 

Multiple UE can collaborate in many ways.
First, multiple UE can quickly establish the global radio feature map of a large ROI.
The local radio feature maps obtained by SLAM from different kinds of UE are subsets of the radio features in the ROI. 
A global radio feature map can be built by fusing the local maps, and the labor-intensive data collection is circumvented.
Second, cooperation helps UE to refine the overlapping parts of the radio feature maps. This approach requires the data association technique in SLAM. 
Although the radio features of each local map may have low confidence, fusing multi-user data can reduce the uncertainty of shared radio features in the global map.
Third, UE with poor sensing ability can directly inherit reliable radio features. For a newly accessed UE, the SLAM algorithm can be initialized by downloading the selected parts of the global map to reduce sensing overhead. 
Each UE continuously refines the legacy radio features through SLAM, adds new radio features, and reports the local map to the cloud database. The cloud refines and distributes the shared features to the corresponding UE.

One of the main challenges of multi-user collaboration is the feature matching of local radio maps in different coordinate systems, where computer vision techniques may be helpful. Data reliability and privacy protection must also be considered. Moreover, avoiding malicious attacks is one of the critical issues for multi-user cooperation. Distributed and centralized cooperation mechanisms must be deeply studied to  utilize computing resources efficiently and save wireless fronthaul and backhaul overhead, in which the advantages of edge computing should be considered.

\subsubsection{Cross-Frequency SLAM}
Cellular systems are expected to ensure communication coverage through low-frequency bands (sub-6 GHz) and high-quality access in hotspot areas via high-frequency bands (mmWave and THz) \cite{co2}. The sub-6 GHz frequency band offers wide coverage and reliable link quality but limited sensing capabilities. High-frequency bands have large bandwidths and antenna arrays, thus enabling high time and angular resolution. However, severe path loss leads to poor communication service quality. Existing works rarely consider cross-frequency SLAM techniques.

Given that 5G NR supports high- and low-frequency dual connections for collocated and distributed deployment, cross-frequency SLAM can form two implementation mechanisms accordingly.
For collocated deployments, the high- and low-frequency bands share some common scatterers.
However, the propagation phenomena through the shared scatterers are different. SLAM results can be represented in different radio map layers with different resolutions.
For example, mmWave/THz transceivers can observe a single specular path associated with a VA, whereas sub-6 GHz transceivers obtain a batch of clustered paths. The mean and spread of the AOA, AOD, and TOA of the clustered paths can be extracted from the received signal in sub-6 GHz, providing rough initial values for high-frequency bands to reduce the sensing overhead. 
For distributed deployments, mmWave/THz transceivers build local radio maps in the local hotspot areas. Sub-6 GHz transceivers are responsible for network control and radio map upload/download. The sub-6 GHz BS can also generate a global radio map by fusing local maps.
SLAM is realized with mmWave measurements, and SLAM results are fused and distributed by sub-6 GHz links.
For example, the sub-6 GHz BS quickly establishes a communication link with the UE and distributes the related local radio features to the UE when a UE enters the service area of a heterogeneous network. 
The UE can quickly access the local mmWave BS by matching its rough position obtained from the GNSS with the downloaded radio features. 
If the UE successfully accesses the mmWave BS, then the subsequent radio map upload/download can be completed through mmWave links.

Many problems must be solved urgently in cross-frequency SLAM techniques. The impact of network switching and execution time on SLAM performance requires further study. The trade-off between the power consumption and SLAM performance of heterogeneous networks must be considered. Wireless network scheduling is an important problem involving RA and optimization for communication and sensing.

\subsubsection{Cross-Device SLAM}
Apart from the possible cooperation of multiple sensing types, users, and frequency bands, interaction among future cellular networks and various devices, such as GNSS, camera, LiDAR, and IMU, exists. These devices typically provide positioning or SLAM services independently. Each has distinct advantages and limitations.

Collaboration across devices can take advantage of different devices and break down limitations. GNSS can provide rough initial values for the relative distance between BS and UE, and then the radio SLAM can work. Cameras can provide rich information about the environment, allowing for robust and accurate place recognition. Common features should be identified for visual and radio SLAM.
This process is crucial to data fusion, because the amount of data and the computational complexity of visual SLAM are unbearable, particularly for IoT devices with limited computation and storage capabilities. UE can access public visual SLAM databases through cellular or Wi-Fi data connections. Thus, complex processing can be performed in the cloud. Moreover, radio SLAM can take the advantage of advanced techniques in visual SLAM. Beamspace channels can be represented as images during the beam sweeping and channel estimation processes. Therefore, deep-learning-based image processing techniques can be applied to extract angle, delay, and Doppler measurements from beamspace channels.
The target surface properties and the propagation environment can be presented as point clouds by active sensing with communication systems. Thus, radio SLAM can be achieved by classifying and semantically segmenting point clouds using graph neural networks.
The IMU integrated into the smartphone can provide the core information of the UE's movement. However, the drift phenomenon is obvious.
The short-term motion measurement is accurate when IMU measurements are incorporated into the motion prediction module of the SLAM algorithm.
Moreover, the drift is corrected in time by radio measurements.

Handling SLAM services across different devices remain challenging on the technical level.
In particular, connections across devices should be preconfigured. This process often requires compatible software and platforms.
In addition, plug-and-play data sharing mechanisms for SLAM should be developed when devices form cross-connections.
Therefore, we may opt in or out of cross-device setups depending on data reliability or situation.

\subsection{SLAM-aided Communications}\label{SLAM-ISAC}
Multi-domain cooperative SLAM makes the communication systems obtain situational awareness. Situational awareness is a concept that goes beyond location awareness. In this concept, the location information of radio nodes and the knowledge of their propagation environments are collected by SLAM techniques. Therefore, the traditional communication blocks  change with the involvement of situational awareness during signal transmission and reception, as shown in Fig. \ref{BD}.

 \begin{figure*}
	\vspace{-0.5cm}
	\centering
	\includegraphics[scale = 0.425]{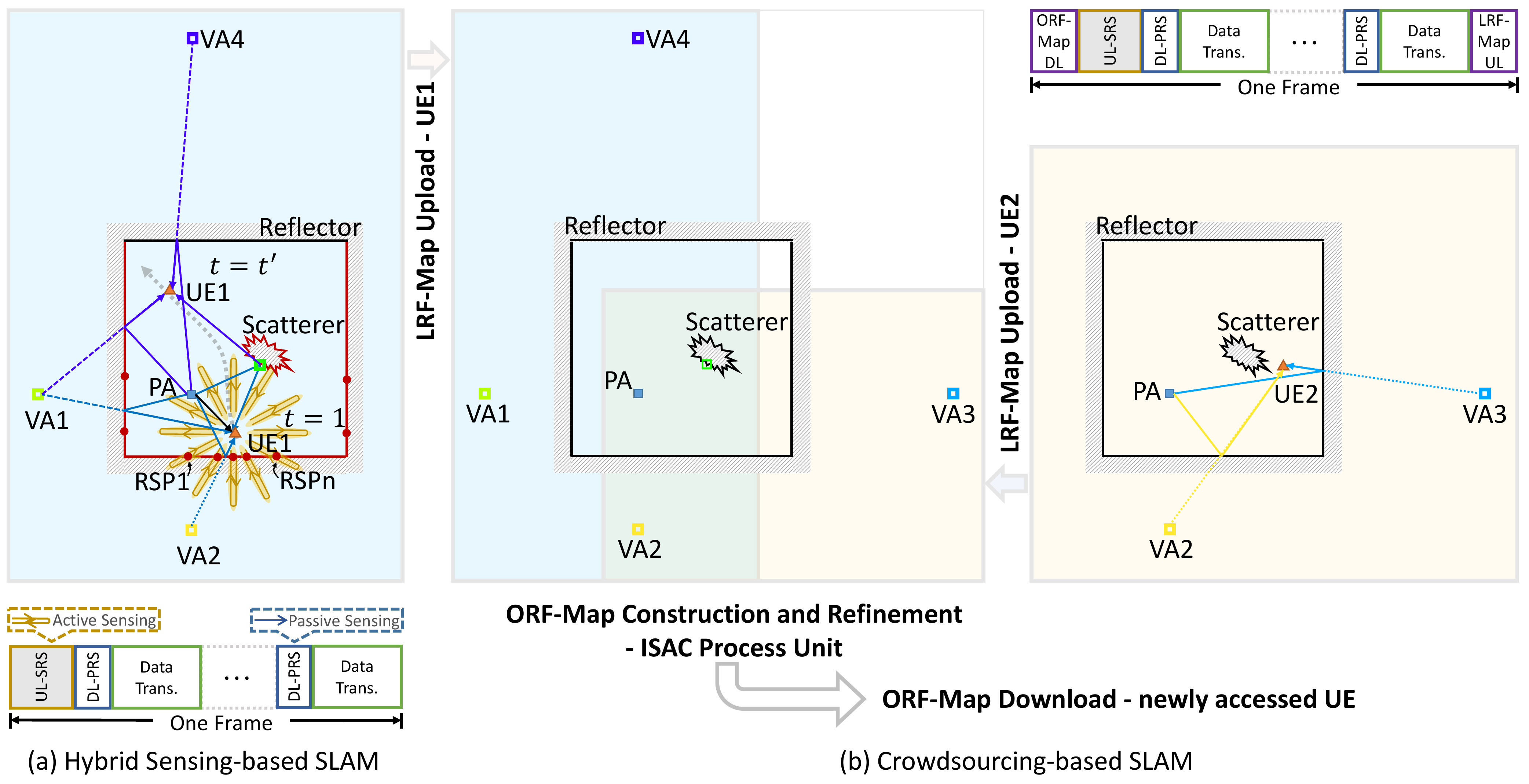}
	\caption{Case studies of the proposed SLAM mechanisms. (a) Hybrid sensing-based SLAM. (b) Crowdsourcing-based SLAM. }
	\label{SLAM}\vspace{-0.2cm}
\end{figure*}

\subsubsection{Beam Management}\label{C-1}
Supporting ultrafast and high-rate data exchanges among moving UE and BSs can hardly be accomplished in traditional mmWave communications because of the unacceptable beam training overhead. Being aware of the reflecting path allows quickly establishing another side link during the blockage. The reflecting path, regarded as a signal from the virtual BS, is obtained by mirroring the BS on the reflecting surface. Although the reflection point is moving when the UE is moving, the virtual BS is static during the UE movement.
Thus, the virtual BS serves as a reference point for beam prediction. The predicted beam directions can greatly narrow down the beam search area, which considerably accelerates the beam tracking process with low overhead. We can quickly determine the beamforming gain that each reflector (or candidate beam) can achieve. Thus, the beam alignment performance at any location in the environment can be foreseen to achieve uninterrupted wireless access.

\subsubsection{Resource Allocation}
Common sensing services, such as positioning or SLAM, can be implemented using communication pilots without consuming additional resources. For high sensing resolution services, such as imaging, sensing reference signals can be allocated through dense comb patterns and periods. The RF chain resources should be allocated for sensing that requires simultaneous transmission and reception, such as active sensing. Therefore, resources should be allocated according to the service requirements of the communication and sensing functions.  Given that SLAM can track channel parameters, such as RSS, AOA, AOD, TOA, and FOA, structured multipath channels can be constructed with these parameters. Therefore, the channel capacity and outage probability can be predicted by using spatial movement coherence combined with channel predictions. Then, additional spectrum resources can be pre-allocated to support unexpected surges in communication traffic demand. Temporary BSs can also be deployed to provide communication links in unfavorable signal outage zones.

\subsubsection{Signal Processing}
Sensing results contribute to the SP strategy design in channel estimation, small-scale channel prediction, CSI feedback, and synchronization. First, SLAM results can be used to refine channel parameters with strong geometric features, such as AOA, AOD, delay, and Doppler, thereby improving channel estimation and prediction performance. In addition, coarse CSI can be predicted from the side information of the UE location and propagation environment. Then, a few pilots are used to estimate the instantaneous small-scale CSI. Thus, the channel estimation overhead is reduced. Adaptive combination mechanisms are required to make instantaneous CSI and environment information complement each other. CSI feedback can also be transformed from traditional mechanisms to situational awareness mechanisms to reduce latency and overhead. Location information can determine a rough synchronization window, simplifying synchronization search operations.

Situational awareness greatly benefits communications by reducing latency and feedback overhead, improving link reliability, and maintaining high throughput. However, a flexible frame structure should be formed to allocate time and frequency resources for sensing and communications. The trade-off between sensing and communication should be further investigated.

\section{Case Studies and Numerical Results}

\subsection{Hybrid Sensing-based SLAM}\label{active_passive_c}
\begin{figure*}
	\vspace{-0.6cm}
	\centering
	\includegraphics[scale = 0.4]{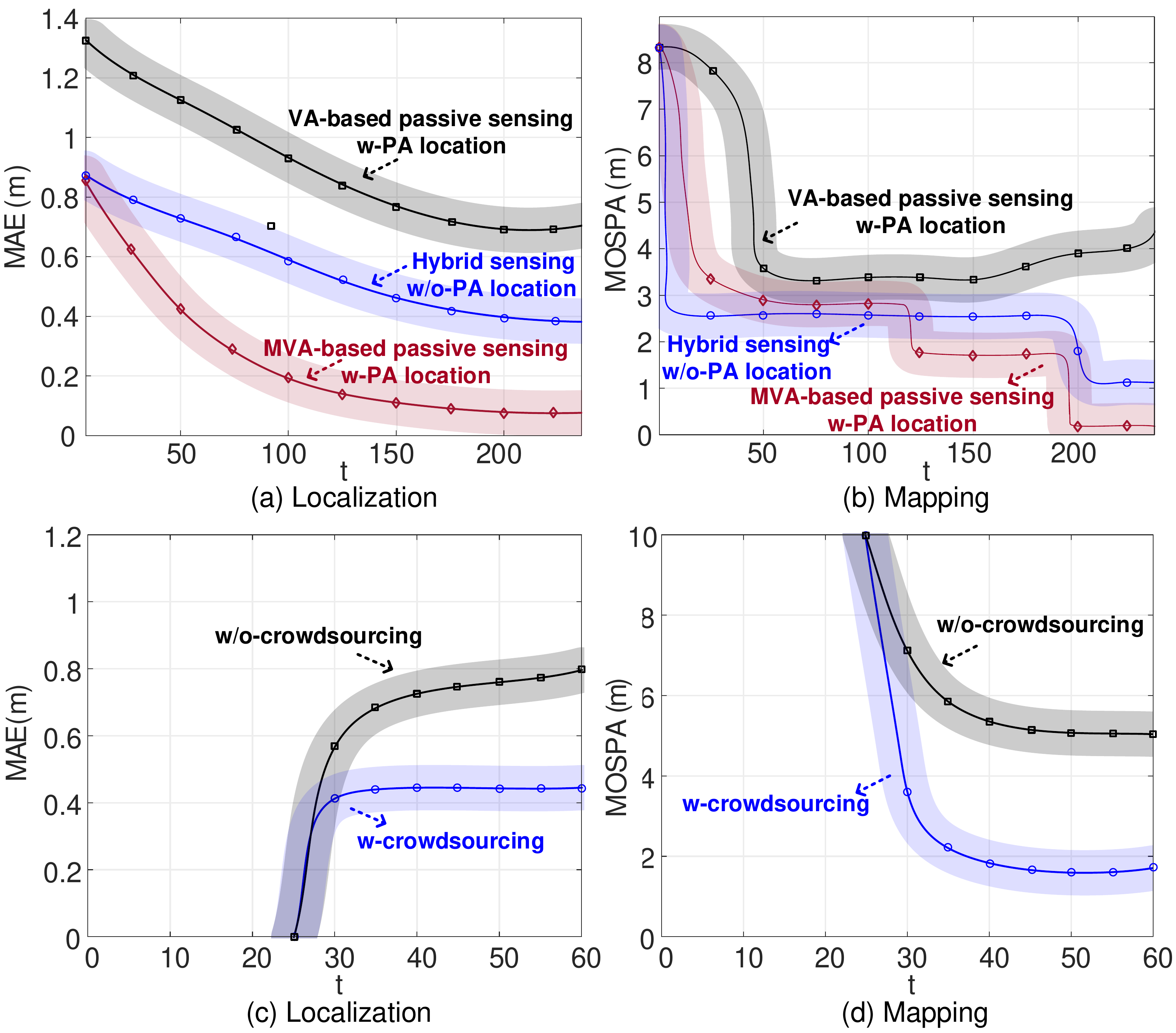}
	\caption{Localization and mapping performance of the proposed SLAM mechanisms. (a) and (b) for hybrid sensing-based SLAM. (c) and (d) for crowdsourcing-based SLAM. }
	\label{SIM}\vspace{-0.2cm}
\end{figure*}

According to Section \ref{active_passive}, an example of the hybrid sensing-based SLAM is depicted in Fig.~\ref{SLAM}(a). Active sensing is performed when the UE sends SRS for initial access while receiving the echo signal. We assume that transmission and reception antenna arrays are placed separately on the UE to relax the self-interference. The corresponding reflective surface point (RSP) can be obtained with the beam direction and the round-trip time. At least two RSPs can determine a reflective surface. A VA can also represent a reflective surface for a specular reflector. Thus, we can convert RSPs (the result of active sensing) into VAs \cite{hap}. 
The DL-PRS for passive sensing that passes through the specular NLOS path can be considered coming from the VA. Therefore, VA becomes the link between active and passive sensing; that is, active and passive sensing can be fused in the same geometric-feature-based radio map. 
Then, we establish an uncertainty model of the results obtained by active sensing to provide the mean and variance of the estimated VAs for soft information fusion with passive sensing.
Next, we extend the classic belief-propagation-based SLAM algorithm by realizing PA initialization with the assistance of active sensing and achieving VA and PA refinement with the help of passive sensing.
Figs.~\ref{SIM}(a) and \ref{SIM}(b) show the localization and mapping performance of the proposed hybrid sensing-based SLAM mechanism compared with those of different passive sensing-only SLAM mechanisms. PA locations are perfectly known for two passive sensing-only SLAM mechanisms (the mechanisms based on VA and master VA [MVA]) \cite{slam}. PA locations are unnecessary in the proposed hybrid sensing-based SLAM mechanism \cite{hap}. We select the mean absolute error (MAE) and mean optimal subpattern assignment (MOSPA) error to measure localization and mapping performance, respectively. The simulation results demonstrate that the proposed mechanism obtains the maximum convergence speed of mapping. Although the proposed mechanism presents a certain performance loss compared with the MVA-based mechanism, the nonnecessity of prior information in PAs expands the application scenarios of the proposed mechanism. 

\subsection{Crowdsourcing-based SLAM}\label{crowdsourcing}
The proposed crowdsourcing-based SLAM mechanism is illustrated in Fig.~\ref{SLAM}(b).
Different kinds of UE cooperate (Section \ref{cu}) to construct and refine the radio map in a decentralized manner. We define the mapping result obtained by each UE as the local radio feature map (LRF-Map). Moreover, we define a set in the cloud as an open radio feature map (ORF-Map), which contains radio features in the ROI. The ORF-Map is a dynamic set that continuously receives information from UE for updates. The frame structure of each UE is illustrated in Fig.~\ref{SLAM}(b). At the very beginning, the ORF-Map is empty. SLAM is performed on the UE side, and the LRF-Maps obtained from multiple UE are subsets of the radio features (PA, VA, and scatterers) in the ROI. Each UE uploads its LRF-Map to the ISAC processing unit at the network center or cloud. Then, the ORF-Map is constructed by fusing algorithms. Therefore, labor-intensive data collection is avoided in the crowdsourcing-based SLAM mechanism. Although the radio features of each LRF-Map may present low confidence levels, common radio features (VA 2 in Fig.~\ref{SLAM}) can be refined with improved reliability via crowdsourcing data fusion. The established ORF-Map can be downloaded with various applications. The downloaded information can be
used to complete the LRF-Maps of the already-accessed UE. For newly accessed UE, the ORF-Map is downloaded at the first moment according to the frame structure, thereby providing good initial values. In particular, the ORF-Map can provide candidate legacy features to each UE in need \cite{coj}. Each UE continuously refines the LRF-Map (by adding new  features, checking the existence of legacy features, and deleting unreliable and vanished features) and reports the LRF-Map to the ISAC processing unit. Figs.~\ref{SIM}(c) and \ref{SIM}(d) show the localization and mapping performance of the proposed crowdsourcing-based SLAM mechanism. We consider eight users with the combination of AOA and TOA measurements, where the orientation and clock biases are considered. The entering time is $1$ for UE 1, 2, and 3; $5$ for UE 4; $10$ for UE 5; $15$ for UE 6; $20$ for UE 7; $25$ for UE 8. The performance of MAE and MOSPA of UE 8 is presented in Figs.~\ref{SIM}(c) and (d), respectively. Compared with the non crowdsourcing mechanism, the crowdsourcing-based SLAM mechanism exhibits improved average positioning and mapping accuracy of UE 8 by $42.5\%$ and $64\%$, respectively, at time $60$. The simulation result verifies the effectiveness of the proposed crowdsourcing-based SLAM mechanism.

The above two case studies investigated the multi-domain cooperative SLAM performance in communication networks.
	We also investigate the SLAM-aided beam tracking to show the performance of SLAM-aided communications, as described in Section \ref{C-1}.
	The designed beam tracking module utilizes the prior information generated by the SLAM algorithm and IMU to narrow down the beam searching scale.
	A switching module is established to monitor the phenomenon of beam birth and death.
	This module enables the system to switch flexibly between the full-scale beam sweeping and the small-scale beam tracking modules.
	Simulations verify that the proposed SLAM-aided beam tracking mechanism can reduce the beam training overhead in complex wireless propagation environments and achieve submeter-level localization and mapping accuracy.

\begin{figure*}
	\vspace{-0.5cm}
	\centering
	\includegraphics[scale = 0.9 ]{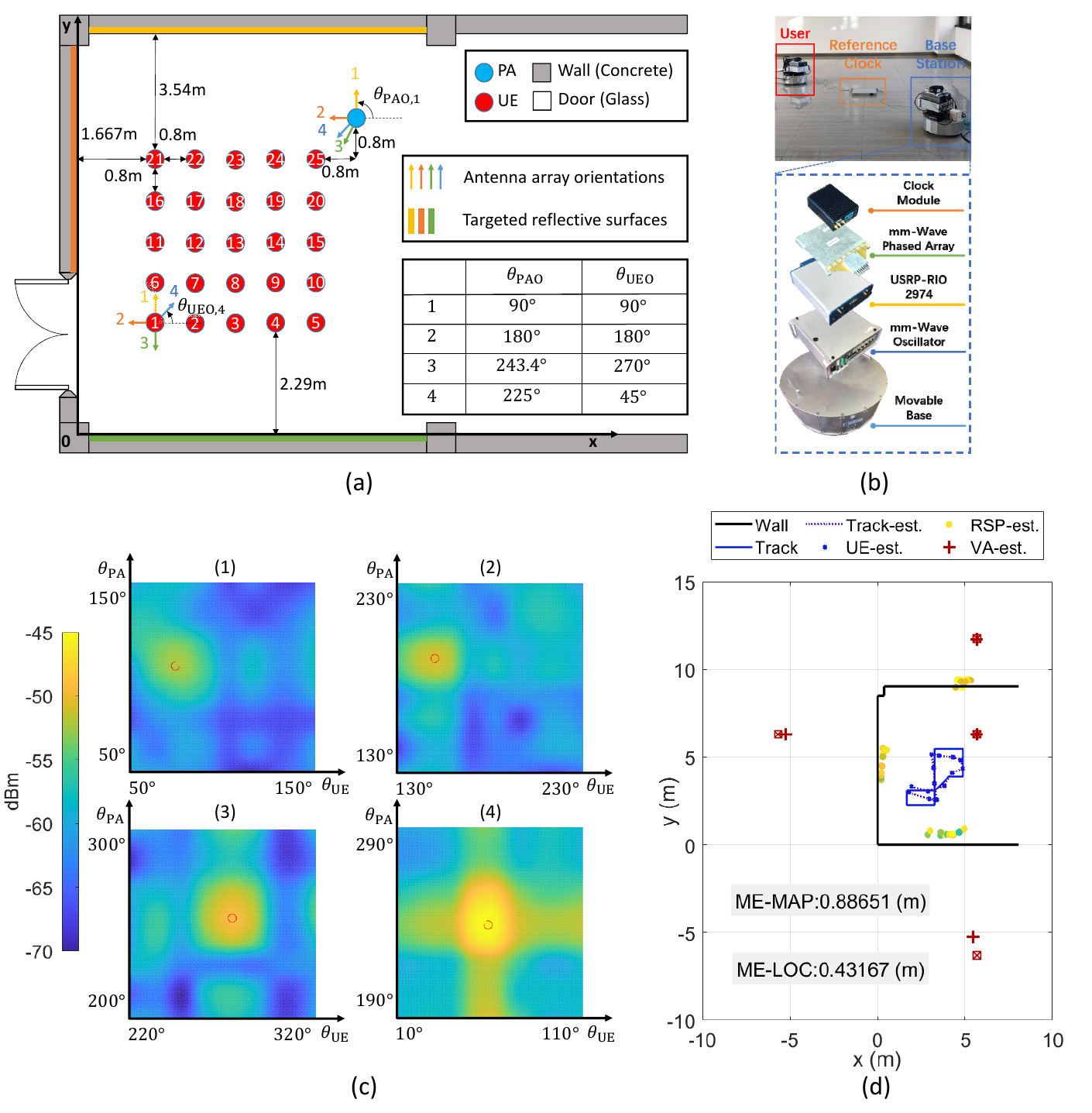}
	\caption{(a) Floor plan of the experimental scenario. (b) Hardware components. (c) RSRP measurements. (d) Experimental results of the proposed SLAM mechanism.}
	\label{RS}\vspace{-0.2cm}
\end{figure*}

\section{Experimental Results}\label{prototype}

We conduct proof-of-concept experiments over the air to validate the feasibility of the proposed SLAM mechanisms. 
The experiment scenario is shown in Fig. \ref{RS}(a). The PA is located at $(5.667, 6.290)$ m. We mark the UE's possible locations with $25$ grid points to describe its track. 
Other complicated scenarios, including those with corridors, multiple wooden doors, decorative glass walls, and pillars, are also considered.
In these scenarios, the phenomenon of beam birth and death exists.
We do not discuss them in this paper because of the page limitation.

The UE and the PA perform beam sweeping according to 5G NR BM.
Our prototype system is composed of the mmWave phased array (mmPSA-1808),  the SDR (USRP-RIO 2974), the mmWave local oscillator, the reference clock node (WR LEN), and the reference clock source (WRS3-18), as shown in Fig. \ref{RS}(b). The working frequency range is $27\!-\!29$ GHz, the center frequency is $28$ GHz, and the transceiver duplex mode is TDD. The PA and the UE share a basic $10$ MHz reference clock source. The USRP-RIO supports a bandwidth of $160$ MHz. The $2.8$ GHz intermediate frequency signal is upconverted to $28$ GHz through the local oscillator. The control signal of the mmWave phased array is generated by the GPIO port of the USRP-RIO to realize beam switching in strict accordance with the corresponding 5G NR synchronization signal block resource.
The mmWave phased array is a linear array with eight antenna elements and one RF chain.
We design eight beam patterns, with eight narrow beams covering $100^{\circ}$ of the space. 
We adopt the bidirectional exhaustive beam sweeping method.
Then, $64$ pairs of beam RSRPs are measured within $4$ ms at the UE side.
In traditional communication systems, the UE compares $64$ RSRP measurements to find the optimal transmitted and received beam pair.
The UE feeds this pair back to the PA in the corresponding time slot through the initial access process.
We set four array orientations of the UE and PA (Fig. \ref{RS}[a]) to map the surrounding reflective surfaces.
The beam RSRP is shown in Fig. \ref{RS}(c), where (1) to (4) correspond to four array orientations. Therefore, we obtain $256$ pairs of beam RSRPs. We extract the AOA and AOD measurements from the beam RSRPs via the successive cancellation method. 
Then, we obtain the SLAM results by conducting the belief-propagation-based SLAM algorithm with AOA and AOD measurements (Fig. \ref{RS}[d]). The solid blue line is the true track, the dotted blue line is the estimated track, the red cross is the estimated VA, and the dot in the heat map color is the estimated scatterer. 
ME-MAP and ME-LOC denote the mean errors of mapping and localization, respectively.
The results verify that the electromagnetic wave at $28$ GHz experiences specular reflection at the smooth wall. Thus, mechanisms that use geometric-feature-based radio maps for SLAM are feasible. The experimental results also show that SLAM can be realized with communication signals and during the beam sweeping process. Moreover, the proposed SLAM mechanism can achieve decimeter-level localization and mapping accuracy.


\section{Conclusion}
In this article, we first summarized the differences between ISAC and traditional communication systems in the aspects of network architectures, protocols, ISAC processing unit,  baseband SP, and RF transceivers.
SLAM can provide the locations of UE and radio features in the propagation environment.
These results help design sensing-aided communication strategies.
We presented key ideas for radio maps, cross-sensing, cross-user, cross-frequency, and cross-device SLAM mechanisms by focusing on multi-domain cooperation. Research opportunities and challenges were also presented. 
In addition, SLAM-aided communications involving BM, RA, and SP were discussed and outlined.
Credible case studies, including hybrid sensing- and crowdsourcing-based SLAM mechanisms, effectively improved the accuracy of localization and mapping.
Finally, the experimental results verified that the proposed SLAM mechanisms can achieve decimeter-level localization and mapping accuracy in practical scenarios.
Therefore, the multi-domain cooperative SLAM is essential in achieving deep integration of sensing and communications.

\section{Acknowledgments}
This work was supported in part by the National Natural Science Foundation of China (NSFC) under Grants 61921004, 61941104, the Key Research and Development Program of Shandong Province under Grant 2020CXGC010108, and 
the Fundamental Research Funds for the Central Universities 2242022k30005.
The work of C.-K. Wen was supported in part by Qualcomm through a Taiwan  University Research Collaboration Project.

\begin{IEEEbiographynophoto}{Jie Yang} \setlength{\baselineskip}{10pt}
	[S'18, M'22] (yangjie@seu.edu.cn)
	received the B.S. degree in communication engineering from Nanjing University of Science and Technology, Nanjing, China, in 2015, the M.S. and Ph.D. degrees in information and communications engineering from Southeast University, Nanjing, China, in 2018 and 2022, respectively. In 2022, she joined the School of Automation, Southeast University, Nanjing, China, where she is currently an Assistant Professor. Her current research interests include signal processing for wireless communications, massive MIMO, millimeter-wave wireless communications, and integrated sensing and communications.
\end{IEEEbiographynophoto}

\begin{IEEEbiographynophoto}{Chao-Kai Wen} \setlength{\baselineskip}{10pt}
	[S'00, M'04, SM'20] (chaokai.wen@mail.nsysu.edu.tw)
	received the Ph.D. degree from the Institute of Communications Engineering, National Tsing Hua University, Taiwan, in 2004.
	He was with Industrial Technology Research Institute, Hsinchu, Taiwan and MediaTek Inc., Hsinchu, Taiwan, from 2004 to 2009.
	Since 2009, he has been with National Sun Yat-sen University, Taiwan, where he is Professor of the Institute of Communications Engineering.
	His research interests center around the optimization in wireless multimedia networks.
\end{IEEEbiographynophoto}

\begin{IEEEbiographynophoto}{Xi Yang} \setlength{\baselineskip}{10pt}
	(yangxi@seu.edu.cn) received the B.S. degree in electrical engineering from University of Electronic Science and Technology of China, Chengdu, China, in 2013, 
	the M.S. and Ph.D. degrees in information and communications engineering from Southeast University, Nanjing, China, in 2016 and 2022, respectively. 
	His research interests center around the signal processing in wireless communications.
\end{IEEEbiographynophoto}

\begin{IEEEbiographynophoto}{Jing Xu} \setlength{\baselineskip}{10pt}
	(xujing\_@seu.edu.cn)
	received the B.S. and M.S. degrees in information and communications engineering from Southeast University, Nanjing, China, in 2019 and 2022, respectively.
	His research interests include integrated sensing and communications.
\end{IEEEbiographynophoto}

\begin{IEEEbiographynophoto}{Tao Du} \setlength{\baselineskip}{10pt}
	(dutao@seu.edu.cn)
	received the B.S. degree in electronic and information engineering from Nanjing University of Science and Technology, Nanjing, China, in 2018, the M.S. degree in information and communications engineering from Southeast University, Nanjing, China, in 2021. He is currently working towards the Ph.D. degree in information and communications engineering with Southeast University, Nanjing, China. His current research interests center around the integrated sensing and communication.
\end{IEEEbiographynophoto}

\begin{IEEEbiographynophoto}{Shi Jin} \setlength{\baselineskip}{10pt}
	[S'06, M'07, SM'17] (jinshi@seu.edu.cn) received his
	Ph.D. degree in communications and information systems from
	Southeast University in 2007. From June 2007 to October 2009,
	he was a research fellow with the Adastral Park Research Campus, University College London, United Kingdom. He is currently with the faculty of the National Mobile Communications
	Research Laboratory, Southeast University. His research interests
	include space-time wireless communications, information theory, intelligent communications, and reconfigurable intelligent
	surfaces. He and his coauthors received the 2010 Young Author
	Best Paper Award from the IEEE Signal Processing Society and
	the 2011 IEEE Communications Society Stephen O. Rice Prize
	Paper Award in the field of communication theory.
\end{IEEEbiographynophoto}

%

%


\begin{thebibliography}{10}
\bibitem{6G0}
A. Bourdoux \textit{et al.}, ``6G white paper on localization and sensing," 2020, [online]. Available: https://arxiv.org/abs/2006.01779.

\bibitem{ISAC00}
F. Liu, Y. Cui, C. Masouros, J. Xu, T. X. Han, Y. C. Eldar, and S. Buzzi,
``Integrated sensing and communications: Towards dual-functional wireless
networks for 6G and beyond," \textit{IEEE J. Sel. Areas Commun.}, pp.1–1, Mar. 2022.

\bibitem{loc}
R. D. Taranto, S. Muppirisetty, R. Raulefs, D. Slock, T. Svensson, and H. Wymeersch, ``Location-aware communications for 5G networks: How location information can improve scalability, latency, and robustness of 5G,'' \emph{IEEE Signal Process. Mag.}, vol. 31, no. 6, pp. 102-112, Nov. 2014.

\bibitem{ISAC2}
J. A. Zhang \textit{et al.}, ``Enabling joint communication and radar sensing in mobile networks-A survey," \textit{IEEE Commun. Surveys \& Tutorials}, vol. 24, no. 1, pp. 306-345,  Jan. 2022.

\bibitem{ISAC11}
A. Liu, Z. Huang, M. Li, et al. ``A survey on fundamental limits of integrated sensing and communication," \textit{IEEE Commun. Surveys \& Tutorials}, vol. 24, no. 2, pp. 994-1034, Feb. 2022.

\bibitem{ISAC1}
D. K. Pin Tan \textit{et al.}, ``Integrated sensing and communication in 6G: Motivations, use cases, requirements, challenges and future directions," in \textit{proc. 1st IEEE International Online Symposium on Joint Communications \& Sensing (JC\&S)}, Mar. 2021, pp. 1-6.

\bibitem{slam}
E. Leitinger, F. Meyer, F. Hlawatsch, K. Witrisal, F. Tufvesson, and M. Z. Win, ``A belief propagation algorithm for multipath-based SLAM,'' \emph{IEEE Trans. Wireless Commun.}, vol. 18, no. 12, pp. 5613-5629, Sept. 2019.

\bibitem{slam2}
R. Mendrzik, F. Meyer, G. Bauch, and M. Z. Win, ``Enabling situational awareness in millimeter wave massive MIMO systems,'' \emph{IEEE J. Sel. Topics Signal Process.}, vol. 13, no. 5, pp. 1196-1211, Sept. 2019.

\bibitem{slam3}
H. Durrant-Whyte and T. Bailey, ``Simultaneous localization and mapping: Part I,"  \emph{IEEE Robotics \& Automation Mag.}, vol. 13, no. 2, pp. 99-110, Jun. 2006.

\bibitem{co0}
M. Z. Win, Y. Shen, and W. Dai, ``A theoretical foundation of network localization and navigation," \textit{Proc. IEEE}, vol. 106, no. 7, pp. 1136-1165, Jul. 2018.

\bibitem{beam}
M. Giordani, M. Polese, A. Roy, D. Castor, and M. Zorzi, ``A Tutorial on Beam Management for 3{GPP NR} at mm{W}ave Frequencies," \textit{IEEE Commun. Surveys Tutorials}, vol. 21, no. 1, pp. 173-196, Sept. 2019.

\bibitem{UAV}
Y. Zeng and X. Xu, ``Toward environment-aware 6G communications via channel knowledge map," \textit{IEEE Wireless Commun.}, vol. 28, no. 3, pp. 84-91, Jun. 2021.

\bibitem{co2} 
N. Gonzalez-Prelcic, A. Ali, V. Va, and R. W. Heath, ``Millimeter-wave communication with out-of-band information," \textit{IEEE Commun. Mag.}, vol. 55, no. 12, pp. 140-146, Dec. 2017.

\bibitem{hap}
J. Yang, C. -K. Wen, and S. Jin, ``Hybrid active and passive sensing for SLAM in wireless communication systems," \textit{IEEE J. Sel. Areas Commun.},  vol. 40, no. 7, pp. 2146-2163, Jul. 2022.

\bibitem{coj}
J. Yang, C. -K. Wen, S. Jin, and X. Li, ``Enabling plug-and-play and crowdsourcing SLAM in wireless communication systems," 
\textit{IEEE Trans. Wireless Commun.}, vol. 21, no. 3, pp. 1453-1468, Mar. 2022.



\end{thebibliography}
\end{document}